\documentclass{edp-conf}
\usepackage{graphicx}
\begin{document}

\TitreGlobal{SF2A 2005}

\title{AMIGA: Very low environment galaxies in the local Universe.}
\author{Verley, S.$^{1,2}$}
\author{Combes, F.}\address{LERMA, Observatoire de Paris, 61, av. de l'Observatoire, 75014 Paris, France}
\author{Verdes-Montenegro, L.}\address{Instituto de Astrof\'i sica de Andaluc\'ia, CSIC, Apdo. 3004, 18080 Granada, Spain}
\author{Leon, S.$^2$}
\author{Odewahn, S.}\address{McDonald Observatory, University of Texas at Austin, Austin TX 78712, USA}
\author{Bergond, G.$^2$}
\author{Espada, D.$^2$}
\author{Garcia, E.$^2$}
\author{Lisenfeld, U.}\address{Universidad de Granada, Granada, Spain}
\author{Sabater, J.$^2$}
\author{Sulentic, J.}\address{Department of Astronomy, University of Alabama, Tuscaloosa, USA}
\runningtitle{Short title here.. }
\setcounter{page}{237}
\index{Author1, A.}
\index{Author2, B.}
\index{Author3, C.}

\maketitle
\begin{abstract} The evolutionary history of  galaxies is thought to be strongly conditioned by the environment. In order to quantify and set limits on the role of nurture one must identify and study an isolated sample of galaxies. But it is not enough to identify a small number of the "most isolated" galaxies. We begin with 950 galaxies from the Catalogue of Isolated Galaxies (Karachentseva 1973) and reevaluate isolation using an automated star-galaxy classification procedure on large digitised POSS-I fields. We define, compare and discuss various criteria to quantify the degree of isolation for these galaxies: Karachentseva's revised criterion, local surface density computations and an estimation of the external tidal force affecting each isolated galaxy. Comparison of multi-wavelength ISM properties, in particular the H$\alpha$ emission line, will allow us to separate the influence of the environment from the one due to the initial conditions at formation.\end{abstract}
%
\section{Introduction}

The role of the environment on galaxy evolution is still not fully understood. In order to quantify and set limits on the role of nurture one must identify and study a sample of isolated galaxies. The AMIGA project, "Analysing the Interstellar Medium of Isolated  Galaxies", is doing a multi-wavelength study of a large sample of isolated galaxies in order to examine their star formation activity and interstellar medium (Verdes-Montenegro et al. 2005, http://www.iaa.csic.es/AMIGA.html).

\section{Isolation study}

Our sample of isolated galaxies gathers 1051 objects, listed in the Catalogue of Isolated Galaxies (CIG - Karachentseva 1973). All of the CIG objects are found in the Catalogue of Galaxies and Clusters of Galaxies (Zwicky et al. 1960-1968; CGCG) with $m_{pg}$ $<$ 15.7 and $\delta >$ -3$^\circ$ ($<$ 3\% of the CGCG). The CIG sample was assembled with the requirement that {\it no similar sized galaxies with diameter d (between 0.25 and 4 times diameter D of the CIG galaxy) lie within 20d}.

Starting from the complete Catalogue, we first remove from our study all galaxies with V $<$ 1500 km s$^{-1}$ (101 CIGs), because the area searched for potential companions of the very nearby CIGs would be extremely large. We reevaluate isolation using an automated star-galaxy classification procedure on large digitised POSS-I fields surrounding each CIG galaxy. The images are reduced using AIMTOOL in LMORPHO (Odewahn 1995), and GUI-driven star-galaxy separation procedure is used to classify detected  sources as: STAR, GALAXY or UNKNOWN.

Then, we define, compare and discuss various criteria to quantify the degree of isolation for these galaxies: e.g. Karachentseva's revised criteria, local surface density computations, estimation of the external tidal force affecting each isolated galaxy. We also apply our pipeline to a subsample of Abell Clusters, Hickson Compact Groups and Triplets of Galaxies which serve as control samples.

\section{H$\alpha$ study}

On 1-2 metre class telescopes, we observed all the CIG spiral galaxies with recession velocities between 1500 km s$^{-1}$ and 5000 km s$^{-1}$, which represent photometric CCD data for a sample of about 200 isolated spiral galaxies.

We took H$\alpha$ and r Gunn images:\newline
(1) H$\alpha$ traces the recent star formation history (last 10.10$^6$ years);\newline
(2) r Gunn defines the bulk of the stellar content.

We focused on the fourth part of them (the bigest and less inclined galaxies) for a detailed morphological study, estimating the angular momentum transfer and therefore the evolution time for bars in isolated galaxies, to be compared with galaxies in denser environments.


\begin{thebibliography}{}
\bibitem{} Casertano, S. and Hut, P. 1985, ApJ, 298, 80
\bibitem{} Dahari, O. 1986, AJ, 89, 966
\bibitem{} Karachentseva, V.~E. 1973, Astrofizicheskie Issledovaniia Izvestiya Spetsial'noj Astrofizicheskoj Observatorii, 8, 3
\bibitem{} Odewhan, S.C. 1995, PASP, 107, 770
\bibitem{} Verdes-Montenegro, L., Sulentic, J., Lisenfeld, U., Leon, S., Espada, D., Garcia, E., Sabater, J., \& Verley, S. 2005, A\&A, 436, 443
\bibitem{} Zwicky, F. and Herzog, E. and Wild, P. 1968, Catalogue of galaxies and of clusters of galaxies
\end{thebibliography}
\end{document}